\documentclass[aps,pra,reprint,showpacs,superscriptaddress,longbibliography]{revtex4-2}
\usepackage{mathtools,amssymb,graphicx,units}

\usepackage[plainpages=false,pdfpagelabels,colorlinks=true,linkcolor=red,urlcolor=blue,citecolor=blue,pdftitle={Title},pdfauthor={},pdfdisplaydoctitle=true,pdfduplex=DuplexFlipLongEdge]{hyperref}
\usepackage{amsmath,graphicx,units}
\usepackage[table,xcdraw]{xcolor}
\usepackage{verbatim}
\usepackage[english]{babel}
\usepackage{color}
\usepackage{ulem}
\usepackage{physics}

\newcommand{\ii}{\mathrm{i}}

\begin{document}
\title{Non-Hermitian Haldane-Hubbard model: Effective description of one- and two-body dissipation}
\author{Can Wang}
\thanks{Equal contribution.}
\author{Tian-Cheng Yi}
\thanks{Equal contribution.}
\author{Jian Li}
\author{Rubem Mondaini}
\email{rmondaini@csrc.ac.cn}
\affiliation{Beijing Computational Science Research Center, Beijing 100193, China}

\begin{abstract}
Using numerically exact diagonalization, we study the correlated Haldane-Hubbard model in the presence of dissipation. Such dissipation can be modeled at short times by the dynamics governed by an effective non-Hermitian Hamiltonian, of which we present a full characterization. If the dissipation corresponds to a two-body loss, the repulsive interaction of the effective Hamiltonian acquires an imaginary component. A competition between the formation of a charge-ordered Mott insulator state and a topological insulator ensues, but with the non-Hermitian contribution aiding in stabilizing the topologically non-trivial regime, delaying the onset of the formation of a local order parameter. Lastly, we analyze the robustness of the ordered phase by following the full dissipative many-body real-time dynamics. An exponentially fast melting of the charge order occurs, whose characteristic rate is roughly independent of the interaction strength, for the case of one-body dissipation.
\end{abstract}

\maketitle

\section{Introduction}
Topological states of quantum systems embody a fundamental departure from the standard classification of spontaneously symmetry-broken phases based on the Landau paradigm of the emergence of a local order parameter. Non-Hermitian Hamiltonians, in turn, also depict a branch off from the standard characterization of certain physical systems~\cite{Ashida2020}, particularly isolated ones, and are significant in describing various processes in nature that involve non-conservation of either energy or particles, including non-equilibrium regimes of open quantum systems~\cite{Ren2022}, optical and photonic systems~\cite{Guo2009, Ruter2010, Feng2011, Regensburger2012, Zeuner2015, El-Ganainy2018, El-Ganainy2019, Miri2019}, mechanical systems~\cite{Liu2016, Yoshida2019a, Brandenbourger2019}, and electric circuits~\cite{Zhang2020, Liu2020, Zou2021, Liu2021, Wu2022}.

A surge of research activity has recently been put forward by mixing these two themes~\cite{Kawabata2019, Bergholtz2021, Okuma2023}, including non-Hermitian versions of the Su-Schrieffer-Heeger model~\cite{Lieu2018, Lee2020, He2021, Ortega-Taberner2022}, models for Chern insulators~\cite{Chen2018, Philip2018, Kawabata2018}, and also models for quantum spin Hall insulators~\cite{Hou2021, Deng2021}. Another fundamental model that encompasses topology is the honeycomb Haldane model~\cite{Haldane1988}, which, by a suitable breaking of time-reversal symmetry, has an associated chiral edge mode mapped by a bulk topological invariant, the Chern number~\cite{Thouless1982}, defining a bulk-boundary correspondence. Its extensions to non-hermiticity exist~\cite{Resendiz-Vazquez2020}, including in modified lattice geometries~\cite{Sarkar2023}. While the standard bulk-boundary correspondence is typically violated under such conditions~\cite{Lee2016, Kawabata2018, Alvarez2018}, it can be recovered with appropriate modifications~\cite{Kunst2018, Yao2018, Yao2018b}.

Focusing on Hermitian scenarios, the Haldane model has been generalized to study the effects of various types of interactions~\cite{Zheng2015,Imriska2016, Shao2021}, which can lead to spontaneous symmetry breaking of the SU(2)-symmetry~\cite{Vanhala2016, Tupitsyn2019, Yuan2023}, or even the emergence of topological characteristics with the increase of interactions for certain electronic fillings~\cite{Mai2023}. Within this correlated setting, the disorder inclusion may lead to a regime of topological Anderson insulator~\cite{Yi2021}, while hopping dimerization can result in higher-order topology~\cite{Yi2023}. Moreover, schemes to drive the real-time dynamics to access target states exhibiting non-trivial topology over fine-tuned time-dependent perturbations have also been developed~\cite{Shao2021b}.

In this paper, we expand the scope of investigations of this paradigmatic model and address the question of the stability of correlated topological phases under dissipative effects, which can be modeled at short times by a non-Hermitian Hamiltonian. We do so by considering either one or two-body dissipation types, which can be fundamentally relevant for understanding the results of the simulations using cold atoms~\cite{Jotzu2014}, where dissipative effects often need to be considered~\cite{Zhou2021, Syassen2008, Tomita2017, Tomita2019, Guardado-Sanchez2021}. Furthermore, this investigation expands the scope of studies of the interplay between topology and interactions in non-Hermitian settings~\cite{Lourenco2018, Yoshida2019b, Yamamoto2019, Yoshida2020, Guo2020, Mu2020, Lee2020, Matsumoto2020, Liu2020b, Zhang2020, Xu2020, Shackleton2020, Zhang2020skin, Xi2021, Lee2021, Yang2021, Yoshida2021, Yamamoto2021, Shen2022, Hyart2022, Yoshida2022, Wang2022, Chen2022, Kawabata2022, Yamamoto2022, Zhang2022, Yoshida2023, Yamamoto2023, Yu2023} for the first time in the case of the honeycomb Haldane model. 

\section{Model and Methods}
Our starting point is the Haldane-Hubbard model in its spinless formulation,
\begin{align}
    \hat H = &-t_1\sum_{\langle i,j\rangle}\left(\hat c_i^\dagger \hat c_j{\phantom{\dagger}} + {\rm H.c.}\right)-t_2\sum_{\langle\langle i,j\rangle\rangle}\left(e^{{\rm i}\phi_{ij}}\hat c_i^\dagger \hat c_j{\phantom{\dagger}} + {\rm H.c.}\right) \nonumber \\
    &+ V\sum_{\langle i,j\rangle}\hat n_i\hat n_j\ + M\sum_{i}(-1)^i \hat n_i,
    \label{eq:H_haldane_hubbard}
\end{align}
where $\hat c_i^\dagger$ ($\hat c_i^{\phantom{\dagger}}$) is the fermionic creation (annihilation) operator in orbital $i$ and $\hat n_i = \hat c_i^\dagger \hat c_i^{\phantom{\dagger}}$ is the corresponding number operator. $t_1$ gives the magnitude of the nearest-neighbor ($\langle i,j \rangle$) hopping terms, whereas $t_2$ is the magnitude of next-nearest neighbor ones ($\langle\langle i,j\rangle\rangle$). The latter further acquires a phase $e^{{\rm i}\phi_{ij}}$ with $\phi_{ij} = +\phi$($-\phi$) for counter-clockwise (clockwise) hoppings [see Fig.\ref{Fig.1}(a)]. Lastly, a staggered potential assigns different onsite energies $\pm M$ for the two sublattices that compose the honeycomb lattice. The noninteracting ($V = 0$) ground state of Eq.~\ref{eq:H_haldane_hubbard} exhibits topological characteristics provided that $M < 3\sqrt{3}t_2\sin \phi$, associated with protected chiral edge modes, mapped by the existence of a topological invariant, the Chern number $C$, related to a physical quantity, the Hall conductivity ($\sigma_{xy} = \frac{e^2}{h}C$)~\cite{Haldane1988, Thouless1982}. Introducing interactions, existing studies have shown that a first-order transition from a topological insulator (TI) to a charge-density wave (CDW) insulator occurs for a critical value of the interactions $V$~\cite{Varney2010, Wang2010, Varney2011}. We aim to understand the nature of this transition under the consideration that the system can be subjected to losses to an environment (bath), which naturally occurs in cold atom experiments via either atom losses or inelastic scattering among them. Under these conditions, the system's density matrix $\hat \rho$ can be described by a quantum master equation,
\begin{align}
    \partial_t\hat \rho(t) = -{\rm i}\left[\hat H, \hat \rho(t)\right] + \hat {\cal L}(\hat \rho)
\label{eq:master}
\end{align}
where $\hat {\cal L}(\hat \rho)$, the Liouvillian superoperator, represents the influence of the bath on the system. Concretely, it reads 
\begin{align}
    \hat {\cal L}(\hat \rho) = -\gamma \sum_m \left(\hat L_m \hat \rho \hat L_m^\dagger -\frac{1}{2}\left(\hat \rho \hat L_m^\dagger\hat L_m + \hat L_m^\dagger\hat L_m\hat \rho\right)\right)\ ,
\end{align}
where $\gamma>0$ describes the loss rate to the environment, assumed homogeneous across the system, and $\hat L_m$, the Lindblad operator in site $m$, denotes the microscopic coupling to the bath. At short times, one can neglect the quantum jump term such that the density matrix approximately evolves as
\begin{equation}
    \partial_t\hat \rho(t) = -{\rm i}\left[\hat H_{\rm eff} \hat \rho(t)- \hat \rho(t)\hat H_{\rm eff}^\dagger\right]\ ,
\end{equation}
where 
\begin{equation}
    \hat H_{\rm eff}=\hat H-{\rm i}\frac{\gamma}{2}\sum_m\hat L_m^\dagger\hat L_m\ .
\end{equation}
This effective non-Hermitian Hamiltonian characterizes the initial dynamics via the analysis of its properties; for example, being non-Hermitian entails that its eigenvalues are generally complex, and the imaginary part governs the corresponding eigenstate lifetime. A single-particle loss to the environment is emulated by $\hat L_m = \hat c_m$, whereas a two-particle inelastic scattering via $\hat L_{mm^\prime} = \hat c_m\hat c_{m^\prime}$. In the first case, $\hat H_{\rm eff}^{(1)} = \hat H - {\rm i} \frac{\gamma_1}{2} \hat N$, where $\hat N =\sum_m \hat c_m^\dagger \hat c_m$ is the total number operator. This second term is a constant due to the particle number conservation and, as a result, does not modify the physical properties of $\hat H$, including the location of the topological phase transition.  Consequently, the single-particle loss is only relevant when the quantum jump term is not neglected in the dynamics.

The two-body loss, on the other hand, leads to an effective non-Hermitian Hamiltonian with the same functional form of the original model, but with interactions acquiring an imaginary component: $V\rightarrow V - {\rm i}\gamma_2/2$ [$\hat H_{\rm eff}^{(2)} = \hat H -{\rm i}\frac{\gamma_2}{2}\sum_{\langle i,j\rangle}\hat n_i\hat n_j$]. Similar analysis has been performed for one-dimensional bosonic~\cite{Xu2020, Zhang2020} or fermionic~\cite{Liu2020b} models exhibiting topological behavior. In what follows, we characterize the low-lying spectrum of $\hat H_{\rm eff}^{(2)}$, where we define the ground state as the one that has the smallest real part. For that, we employ a Krylov-Schur-based diagonalization method~\cite{petsc-user-ref, Hernandez2005} to extract the eigenpairs $E_\alpha, |\alpha_R\rangle$, as well as the corresponding left eigenvectors $|\alpha_L\rangle$, satisfying $\hat H_{\rm eff} |\alpha_R\rangle = E_\alpha|\alpha_R\rangle$ [$\hat H^\dagger_{\rm eff} |\alpha_L\rangle = E^*_\alpha|\alpha_L\rangle$].

Calculations are mainly performed in the highly symmetric $N_s = 24$-site cluster [Fig.~\ref{Fig.1}(a)] in the presence of periodic boundary conditions (PBC) at half-filling. Such geometry satisfies two important criteria: It preserves all point-group symmetries of the thermodynamic limit and possesses the Brillouin zone corner $K$ as a valid momentum point, which allows capturing the concomitant occurrence of a topological-to-trivial transition and the quantum phase transition associated with the emergence of charge ordering in this model~\cite{Varney2010, Varney2011}. With PBCs, one can block-diagonalize $\hat H_{\rm eff}$ in momentum sectors since the added non-Hermitian terms do not break the translation invariance present in the original Hamiltonian~\eqref{eq:H_haldane_hubbard}. We note that for all parameters we investigate, the ground state still resides in the zero-momentum sector with $\gamma_1,\gamma_2\neq0$, and our characterization is constrained to it. Lastly, we take $t_1=1$ as the energy unit and focus the calculations at $t_2=0.2$ and $\phi=\pi/2$ in the absence of staggered potentials ($M =0$). This last choice naturally increases the robustness of the topological phase upon the inclusion of interaction since it avoids facilitating the creation of a trivial charge density-wave pattern that would spontaneously emerge at sufficiently large $V$'s in the thermodynamic limit. Taking $\phi=\pi/2$ also makes the topological phase more robust to the interactions~\cite{Varney2010}.

\section{Results}
We start by describing the formation of charge order in the system governed by the real interactions $V$, associated with the spontaneous symmetry breaking of inversion symmetry. This is done by computing the (zero-momentum, ${\bf k} = 0$) expectation values of the CDW structure factor,
\begin{equation}
    S_{\rm cdw}^{ab} = \frac{1}{N_s}\sum_{i,j}\ _a\langle  (\hat n_i^A - \hat n_i^B)(\hat n_j^A - \hat n_j^B)\rangle_b
\end{equation}
where $\hat n_i^{A,B}$ are the number operators in sublattices $A$ and $B$ of the $i$-th unit cell. Here, we notice the ambiguity in computing expectation values, where one can use either $a,b = L,R$ eigenstates. In doing so, we ensure that $_a\langle \hat O\rangle_b \equiv \frac{\langle \alpha_a|\hat O | \alpha_b\rangle}{\langle \alpha_a | \alpha_b\rangle}$ is correctly normalized. We focus on the right eigenstates in the main text, i.e., $a,b=R$, but Appendix~\ref{app:lr_eig} shows the different possibilities, which give compatible results.

\begin{figure*}[tpb]
\includegraphics[clip,width=0.9\textwidth]{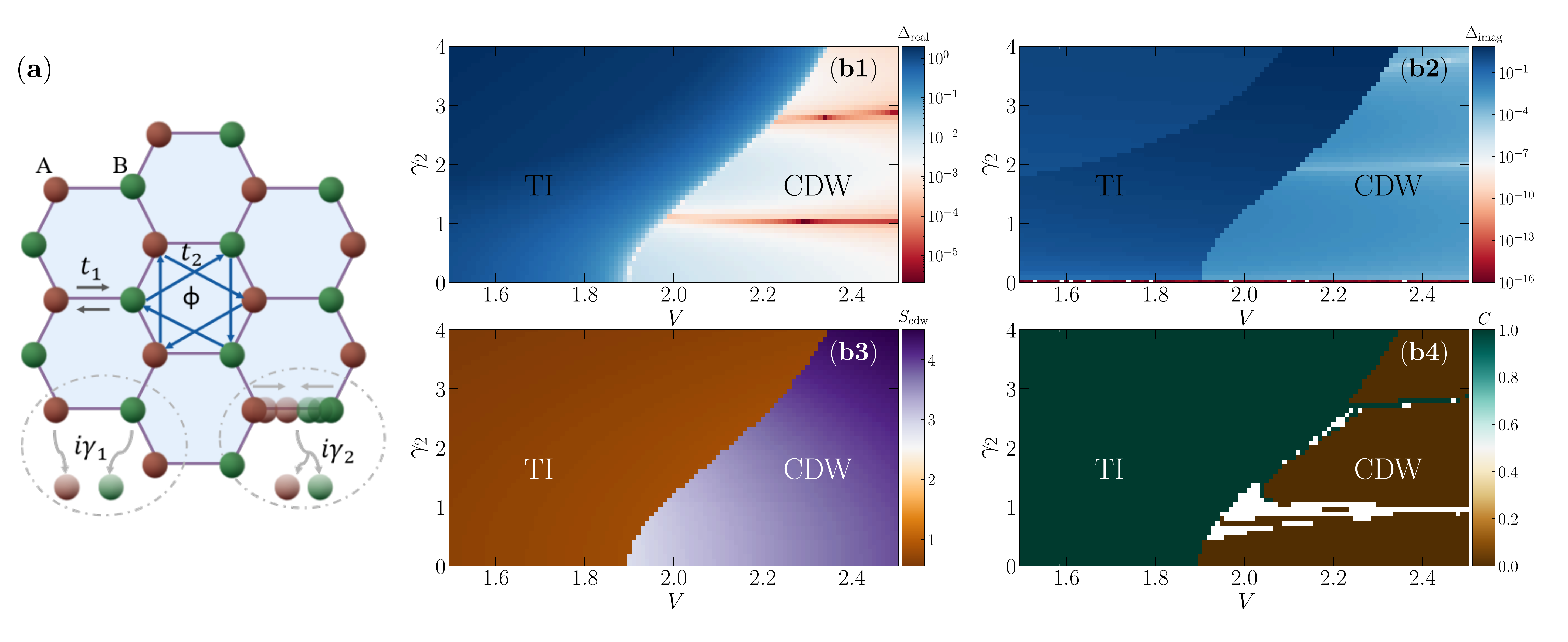}\caption{(a) Schematic diagram of the non-Hermitian Haldane-Hubbard model in the six fold rotationally symmetric lattice with $N_s=24$ sites; the dissipation rates for one- and two-body losses are given by $\gamma_1$ and $\gamma_2$, respectively. (b1) and (b2) show the real parts, $\Delta_{\text{real}}$, and imaginary parts, $\Delta_{\text{imag}}$, of
the excitation gap with two-body dissipation ($\gamma_1 = 0$; $\gamma_2>0$) in the space of parameters $(V,\gamma_2)$; (b3) and (b4) give the corresponding phase diagrams of the CDW structure factor, $S_{\rm cdw}$, and the Chern $C$ under similar conditions. Here,
the parameters are $t_{1}=1,$ $t_{2}=0.2,$ $\phi=\frac{\pi}{2},$ and $\gamma_{1}=0.$}
\label{Fig.1}
\end{figure*}

The results for $S_{\rm cdw}$ in the space of parameters $(V,\gamma_2)$ shown in Fig.~\ref{Fig.1}(b3) highlight that the critical point at which the charge ordering rapidly grows is pushed toward larger interaction strengths $V_c$ once the two-body dissipation is introduced ($\gamma_2>0$). An interpretation is that charge ordering, resulting in a Mott-insulating ground state, is significantly hampered by the loss associated with the interactions; consequently, one would need a larger real $V$ to counterbalance it. Such behavior is also followed by the excitation gap $\Delta = E_1 - E_0$, whose real and imaginary parts are displayed in Fig.~\ref{Fig.1}(b1) and \ref{Fig.1}(b2), respectively. Within the CDW regime, both $\Delta_{\rm imag} \equiv {\rm Im}(E_1-E_0)$ and  $\Delta_{\rm real} \equiv {\rm Re}(E_1-E_0)$ are substantially reduced in comparison to the regime without charge ordering. In the former, the two lowest states in the spectrum form a doublet which are CDW-like states exhibiting even or odd parity under the inversion symmetry and are typically far from the bulk of the spectrum. 

A final characterization of the ground state is made by considering its topological properties. Here, the Chern number
\begin{equation}
C_{ab}=\int \frac{d \phi_{x} d \phi_{y}}{2 \pi {\rm i}}\left(\left\langle\frac{\partial \psi_a}{\partial \phi_x} \biggr\rvert \frac{\partial \psi_b}{\partial \phi_y}\right\rangle-\left\langle\frac{\partial \psi_a}{\partial \phi_y} \biggr\rvert \frac{\partial \psi_b}{\partial \phi_x}\right\rangle\right) \ ,
\label{eq:chern}
\end{equation}
with $a,b=L,R$ is computed by obtaining the many-particle ground-state $|\psi_{a,b}\rangle$ under a set of twisted boundary conditions (TBCs) $\{\phi_x,\phi_y\}$, being quantized for any system size. In principle, this leads to a set of four different Chern numbers, $C_{LL}, C_{LR}, C_{RL}, C_{RR}$, but it has been recently proven that they are all equivalent~\cite{Shen18}, including in many-body settings~\cite{Xu2020}. We choose $C\equiv C_{RR}$ for convenience. Besides that, provided that the excitation gap is finite for all twisted boundary conditions when adiabatically changing a control parameter in the Hamiltonian ($\gamma_2$ or $V$ in our case), the topological number remains invariant. Such gap opening condition~\cite{Fukui2005} is a necessary ingredient for a quantized Chern number, and Eq.~\eqref{eq:chern} can be further computed over a coarsely discretized Brillouin zone defined by the TBCs~\cite{Varney2011, Shao2021}. Here we use a number $N_{\phi_x,\phi_y}=9$ of boundary conditions for each direction.

The corresponding Chern number phase diagram is displayed in Fig.~\ref{Fig.1}(b4). It shows to directly extend the results of the Hermitian case~\cite{Varney2010, Varney2011} in that the topological insulator (TI) survives only in the regime of small interactions and is further benefited from the suppression of the charge ordering by the dissipation rate $\gamma_2$. It thus corroborates that, at half-filling, this model does not support a topological Mott insulator, now extending such results to the non-Hermitian frame.

\begin{figure}[t!]
\includegraphics[clip,width=0.23\textwidth]{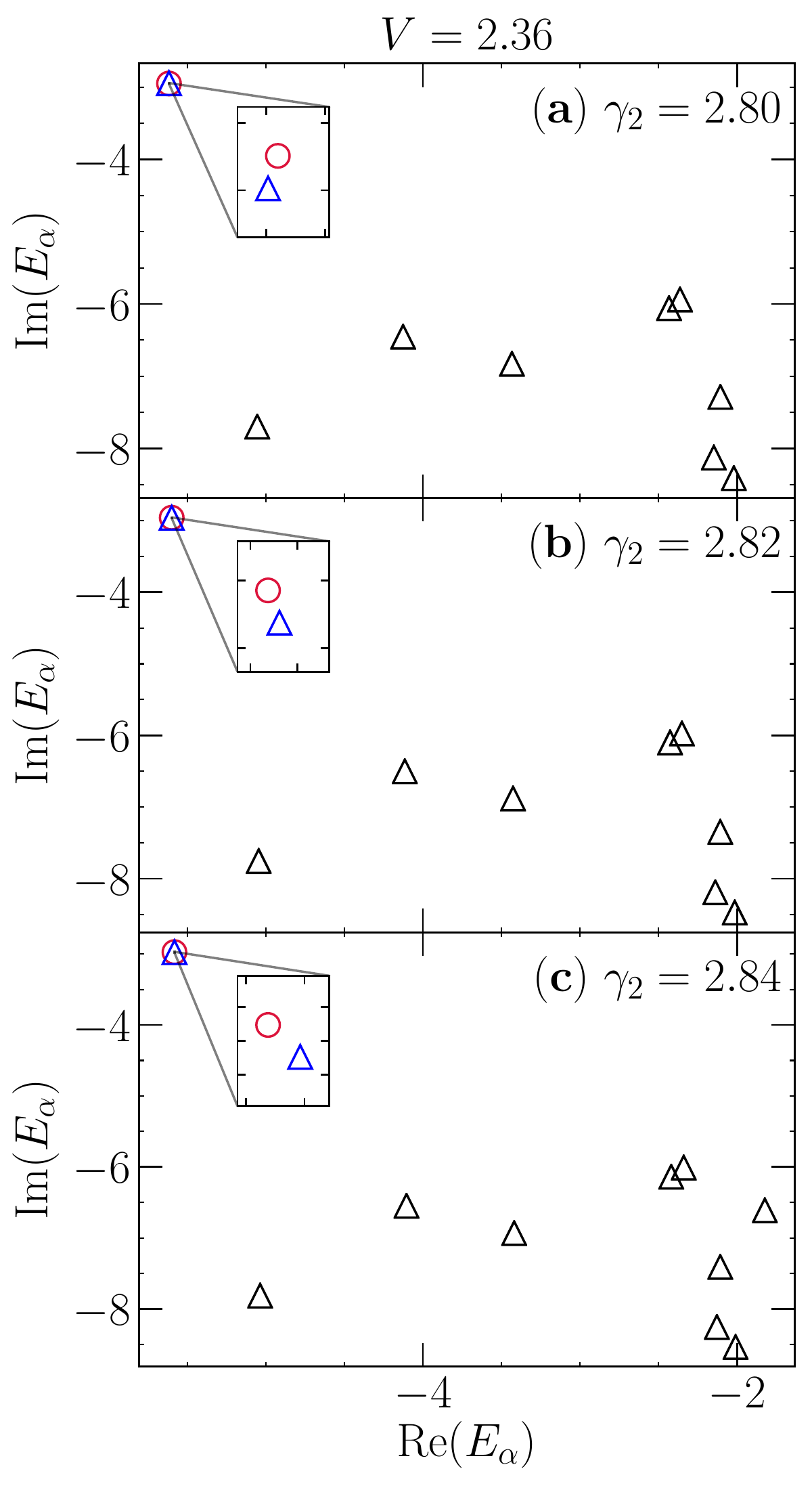}
\includegraphics[clip,width=0.23\textwidth]{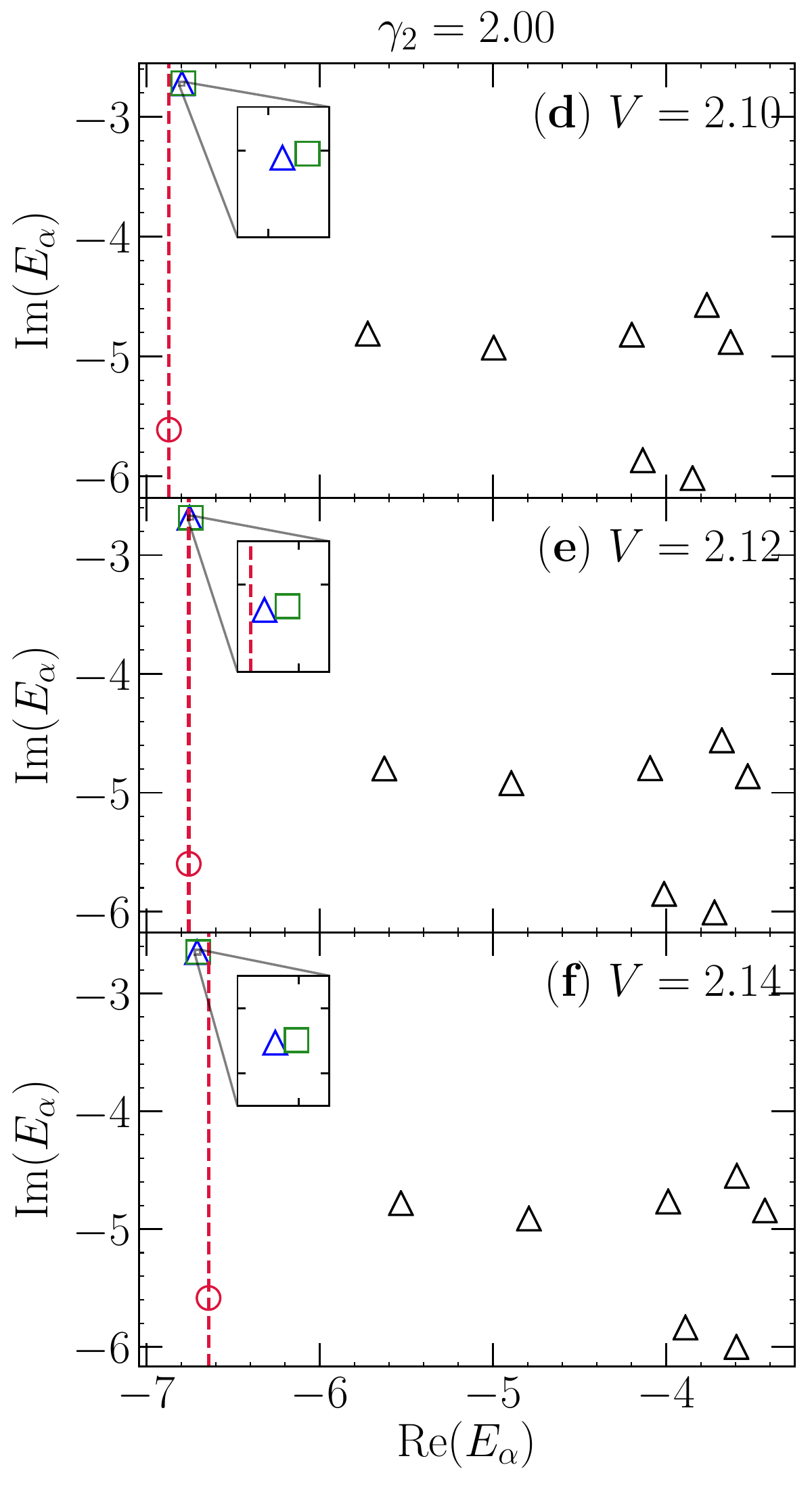}
\caption{The low-lying spectrum for fixed interaction strength $V=2.36$ (a)-(c), and growing $\gamma_2$; it shows that the real part of the eigenenergies of doublet of states cross, establishing that the Chern number gap opening condition is no longer satisfied. (d)-(f) Low-lying spectrum for fixed two-body dissipation rate $\gamma_2=2$ across the TI-CDW phase transition (see text for details). The vertical dashed line provides a visual aid to the location of the ${\rm Re}(E_\alpha)$ for the state that originally exhibits a non-trivial Chern number, helping in visualizing the crossings of its real part with the CDW-like doublet when increasing the interaction strength $V$. In both cases, the zoom-ins focus on the doublet of states which are representative of the CDW phase.}
\label{Fig.2}
\end{figure}

Nonetheless, we note that within the CDW insulator phase, the Chern number can exhibit non-quantized values. A quick comparison to the phase diagram for the gap for excitations [Fig.~\ref{Fig.1}(b1)] suggests that a gap closing for the real part of $\Delta$ is likely tied to such behavior, where branches with vanishing $\Delta_{\rm real}$ appear at close constant values of $\gamma_2$. To understand this better, we report in Fig.~\ref{Fig.2}(a)--\ref{Fig.2}(c) the low-lying spectrum for periodical boundary conditions ($\{\phi_x,\phi_y\}=0$) at a fixed interaction strength ($V=2.36$), near one of such branches ($\gamma_2 =2.80 - 2.84$). The previously mentioned CDW doublet exhibits a level ``crossing'' in that the real part of the gap vanishes while the imaginary part remains finite. That is sufficient to render a non-quantized Chern number. We also note that in Fig.~\ref{Fig.1}(b2), the imaginary part of the gap displays branches in the CDW phase with small values, but these remain finite.

The TI-CDW transition, on the other hand, is characterized by a crossing of the doublet with the original state displaying a finite Chern number, see Figs.~\ref{Fig.2}(d)--\ref{Fig.2}(f), for a finite $\gamma_2=2$. As a result, $\Delta_{\rm real}\to 0$ at $V\to V_c$, but the imaginary part, initially positive, turns negative when the first state in the doublet acquires the smallest real energy and changes again to positive (but small in magnitude) when both states of the doublet have ${\rm Re}(E_\alpha)$ smaller than the original topologically non-trivial state.

\begin{figure}[t!]
\includegraphics[clip,width=0.44\textwidth]{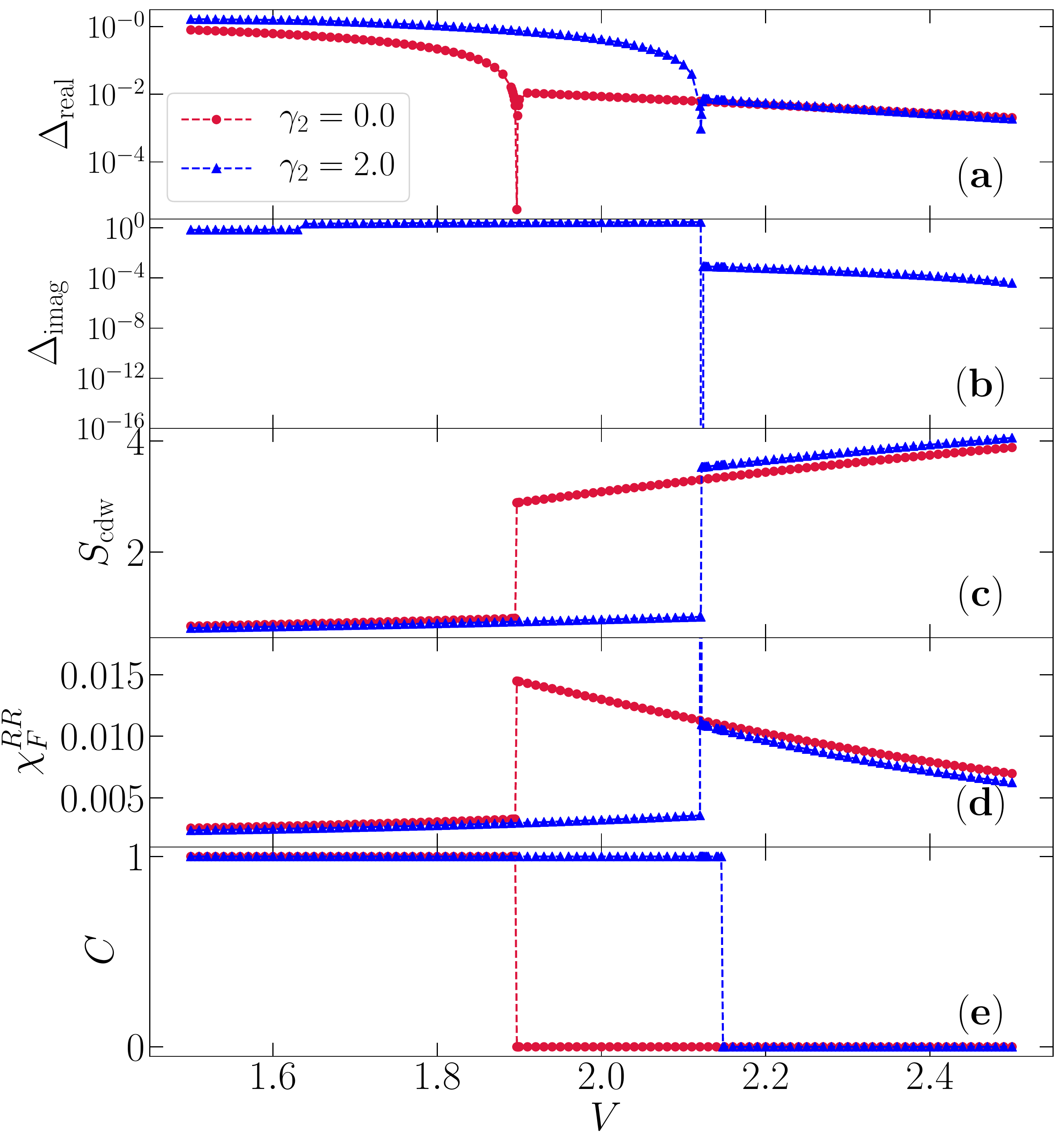}\caption{The real (a) and imaginary (b) parts of the excitation spectrum contrasting the Hermitian ($\gamma_2 = 0$) and non-Hermitian cases ($\gamma_2=2$) as a function of the (real repulsive) interaction $V$. (c) the structure factor $S_{\rm cdw}^{RR}$ (computed in the ground state) dependence in $V$: The discontinuities for $\gamma = 0$ and $>0$ reflect a first-order phase transition in either case. (d) The fidelity susceptibility $\chi^{RR}_F$, quantified with right-eigenvectors and (e), the topological invariant (Chern number) with increasing $V$.}
\label{Fig.3}
\end{figure}

A closer inspection of the type of phase transition is shown in Fig.~\ref{Fig.3}, where we do a line cut in the phase diagram for $\gamma_2 = 0$ and 2. That the topological transition in either Hermitian or non-Hermitian cases coincides with the emergence of charge order can be inferred by the simultaneous drop of the Chern number from one to zero and the discontinuity of the charge structure factor. As previously mentioned, such discontinuity is thus accompanied by the closing of the real part of the excitation spectrum, classifying it thus as a first-order phase transition even when $\gamma_2\neq0$. The imaginary part, while it does not strictly close, also reflects the critical interaction strength $V_c$.

Lastly, further characterization of the phase transition can be obtained by the fidelity susceptibility, a metric that quantifies the dissimilarity between ground states upon a small variation of a control parameter. Here, we compute it via a variation of the real interaction strength $V\to V+dV$, with $dV=10^{-3}$, extracting the fidelity susceptibility as~\cite{Zanardi06, CamposVenuti07, Zanardi07, You2007}
\begin{equation}
    \chi^{ab}_{F} = \frac{2}{N_s}\frac{1 - |\langle \psi_a(V)|\psi_b(V+dV)\rangle|}{dV^2}\ .
\end{equation}
For continuous phase transitions, this quantity exhibits extensive (in the system size) peaks at the critical point~\cite{Yang07, Varney2010, Jia11, Mondaini15, Jin2022}, whereas first-order phase transitions are thus characterized by discontinuities whose amplitude are governed by the small parameter $dV$~\cite{Varney2010}. Agreeing with the analysis of the excitation gaps, we observe the latter behavior in Fig.~\ref{Fig.3}(d), irrespective of whether $\gamma_2$ is finite or not. Since it can be computed with different combinations of left/right eigenstates, we show the case $\chi^{RR}_F$ but notice that similar behavior follows if using other  pairs of ground-states in $\chi^{ab}_F$ (see Appendix~\ref{app:lr_eig}). 

Lastly, we observe that changing the sign of the dissipation $\gamma_2\to -\gamma_2$ leads to the same physical results for all quantities we investigate while resulting in the complex conjugation of the spectrum of the effective non-Hermitian Hamiltonian, $E_\alpha\to E_\alpha^*$. We notice that the complex conjugation of $\hat H_{\rm eff}^{(1,2)}$ itself changes the chirality of the edges modes while physically corresponding to the modification from loss to gain in the dissipation. Such a procedure does not change the loci of the onset of the CDW Mott insulating regime or its first-order characteristics. 

\subsection*{Dynamics in the presence of dissipation}
Having established the main features of the non-Hermitian effective Hamiltonian that emerge when the quantum jump terms are neglected in the quantum master equation, we now fully focus on the actual dynamics under dissipative conditions. Evolution of the full density matrix, as governed by Eq.~\eqref{eq:master} is often prohibitive (even more for the large-Hilbert space size we consider), and an alternative approach is to take the stochastic evolution of quantum trajectories instead~\cite{Dalibard1992, Dum1992, Dum1992b, Plenio1998, Daley2014}. The rationale is that the system evolves under the non-Hermitian Hamiltonian, with the quantum jumps being stochastically applied at certain points in time. In such a procedure, the average of the density matrix (or the corresponding physical observables) for many ``trajectories'' (which depend on the instants where the action of quantum jump terms take place) converges to the one if the full quantum master equation were considered. Since the number of trajectories is relatively small for a satisfactory convergence, the procedure entails a fraction of the computational cost -- details are given in Appendix~\ref{app:qtm}.

\begin{figure}[t!]
\includegraphics[clip,width=0.49\textwidth]{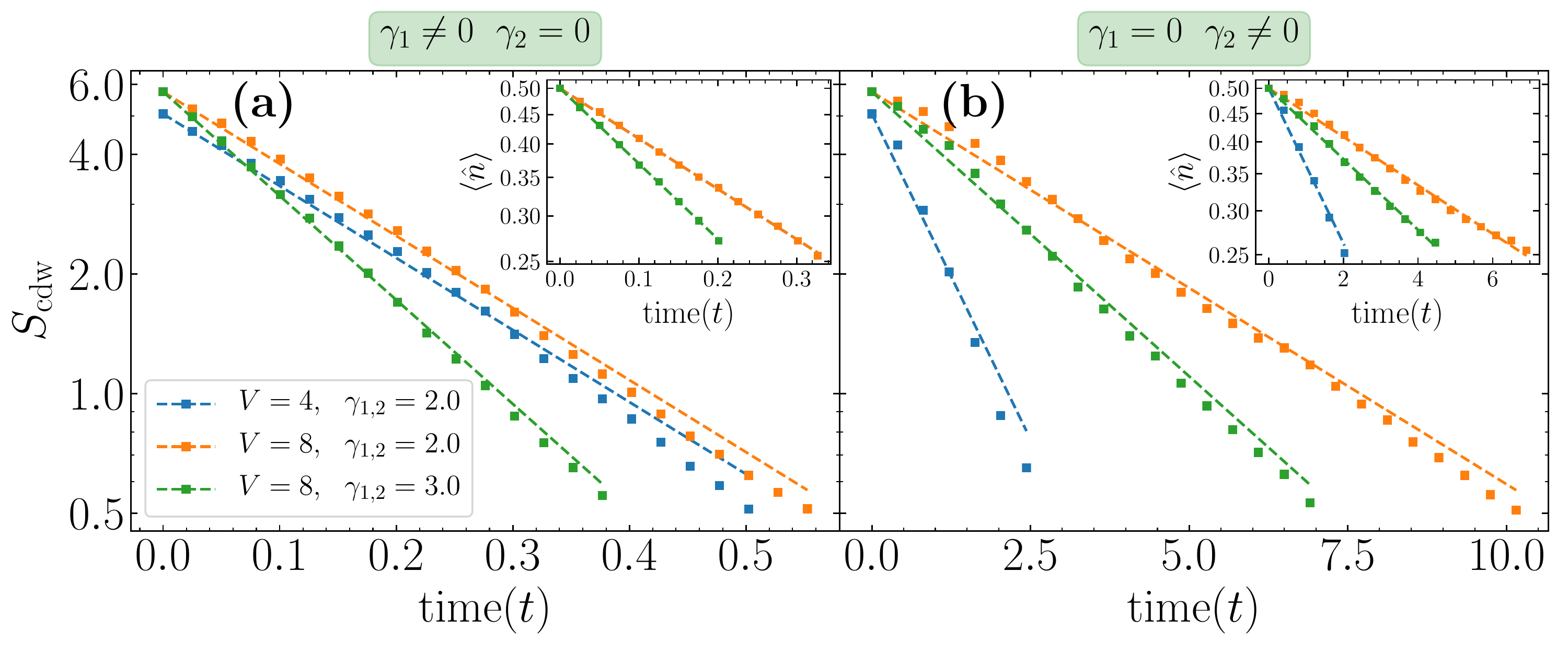}\caption{(a) Dynamics of the CDW structure factor with one-body dissipation. (b) The same for the case with a two-body dissipation. The corresponding insets display the average density time dependence for each case. Dashed lines denote the exponential fitting (note the vertical logarithm scale) for all quantities and regimes of parameters given by the markers. Here, the time scales are in units of the inverse NN hopping energies $\tau = t_1^{-1}=1$, and we work in units that $\hbar=1$.}
\label{Fig.4}%
\end{figure}

Using this procedure, we focus on the dynamics of the CDW structure factor, $S_{\rm cdw}$, observing the melting of charge order under dissipation with either one-body or two-body types. Figure~\ref{Fig.4} reports the results for the time-dependence of $S_{\rm cdw}$ when deep into the Mott insulating phase, i.e., $V \gg V_c(\gamma_{1,2})$, for both $\gamma_1\neq 0$ [Fig.~\ref{Fig.4}(a)] and $\gamma_2\neq 0$ [Fig.~\ref{Fig.4}(b)]. The insets give the corresponding dynamics of the total density $\langle \hat n\rangle=\langle \hat N\rangle/N_s$. Both quantities exhibit an exponential decay in time for various parameter sets but typically display much faster decay in the case of one-body dissipation. In this case, the average density quickly departs from half-filling at time $t=0$, and is rather insensitive to the interaction strength $V$,  while the larger the dissipation rate $\gamma_1$ is, the faster it decays. As a result, the CDW structure factor necessarily decreases from its maximum value at half-filling, denoting the melting of the charge ordering at time scales much shorter than the typical hopping times $\tau = 1/t_1 = 1$. 

\begin{figure}[tpb]
\includegraphics[clip,width=0.49\textwidth]{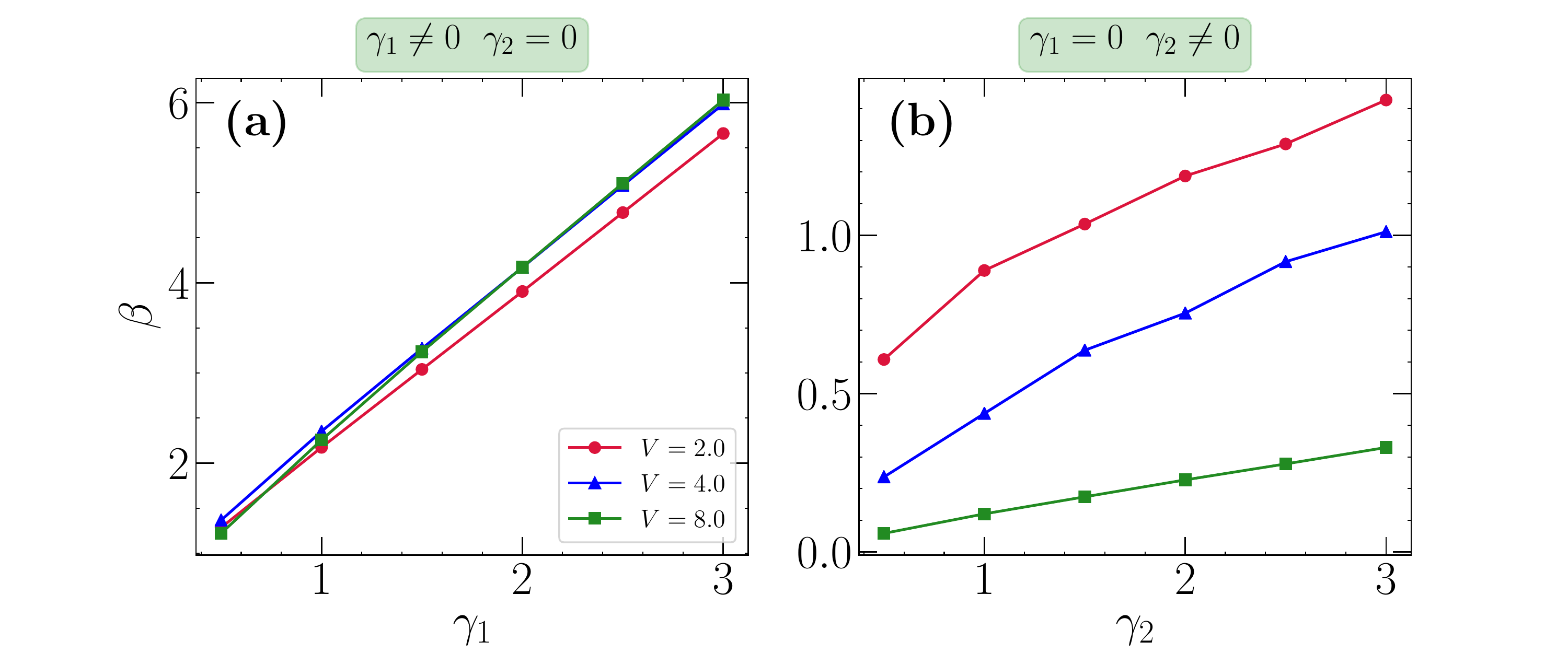}\caption{Decay rate $\beta$ of $S_{\rm cdw}$ with the one-body dissipation $\gamma_1$ (a) and the two-body dissipation $\gamma_2$ (b). While in both cases the larger $\gamma_{1,2}$, the quicker the melting of charge ordering, if focusing on the two-body dissipation, the stability of the Mott phase can be attested at longer time scales.}
\label{Fig.5}
\end{figure}

In turn, while the decay rates are much slower in the two-body dissipation case, decay times of the particle number and $S_{\rm cdw}$ are affected by both the interaction strength $V$ and the dissipation rate $\gamma_2$, but here the deeper one is in the Mott regime, the longer it takes for the melting of charge order. Lastly, by performing an exponential fitting of the CDW structure factor of the form $S_{\rm cdw}\propto \exp(-\beta t)$, we compile in Fig.~\ref{Fig.5} the decay rate $\beta$ for both types of dissipation, as a function of the corresponding dissipation rate $\gamma_{1,2}$, systematically confirming these previous features for an extended parameter set.

\section{Summary and discussion}
We investigate the impact of dissipation on the topologically non-trivial and charge-ordered phases of the Haldane-Hubbard model in its spinless formulation. At short times, the physics is governed by an effective non-Hermitian Hamiltonian, which we comprehensively characterize via the Chern number, CDW structure factor, excitation gaps, and fidelity susceptibility. In this static case, only two-body dissipation exhibits non-trivial behavior, and we notice that it competes with the formation of a charge-ordered Mott insulator, improving the resilience of the topological phase to the presence of the contact interactions $V$.

Whereas these conclusions are born from the study of a non-Hermitian Hamiltonian that neglects the influence of the quantum jump terms in the master equation, the study of the actual dynamics shows that charge ordering is also mitigated with dissipation, but rather on a trivial level: The particle density decreases thus destroying the Mott insulator state and its corresponding CDW order. The fact that the particle density exponentially decays in time also brings challenges for the manifestation of a non-trivial topological regime.

While it has been demonstrated that the topological invariant of a quantum system cannot change under unitary transformations~\cite{Chen10, Alessio15, Caio16}, the presence of dissipation evades this paradigm, allowing one to dynamically change the value of the Chern number, for example. Besides that, if a system is then comprised by the simultaneous presence of gain and loss, as the one introduced in Ref.~\cite{Resendiz-Vazquez2020} for the Haldane model, the average density remains close to its initial value at all times, and the dynamical manipulation of the Chern number becomes likely feasible. We leave such investigation for future studies. 

Lastly, we stress that the emulation of the Haldane model with trapped cold atoms in optical lattices~\cite{Jotzu2014}, combined with techniques that permit the control of contact (non-local) interactions using Rydberg dressing~\cite{Guardado-Sanchez2021}, represent the ideal platform to experimentally study the results we numerically unveil, even more so because atom loss is an inherent effect in these experiments. Similar to what we observe, an exponential decay on the atom number has been reported in Ref.~\cite{Guardado-Sanchez2021} melting a charge-density wave state whose dissipation mechanism is akin to the single-particle loss we investigate.

\section{Acknowledgments}
W.C. thanks Supeng Kou, Gaoyong Sun, and Yang Xue for useful discussion. R.M.~acknowledges support from the NSFC Grants No.~NSAF-U2230402, No.~12111530010, No.~12222401, and No.~11974039. Numerical simulations were conducted in the Tianhe-2JK at the Beijing Computational Science Research Center.

\appendix

\section{Left and right eigenstates} \label{app:lr_eig}
\begin{figure}[htp]
\includegraphics[clip,width=0.45\textwidth]{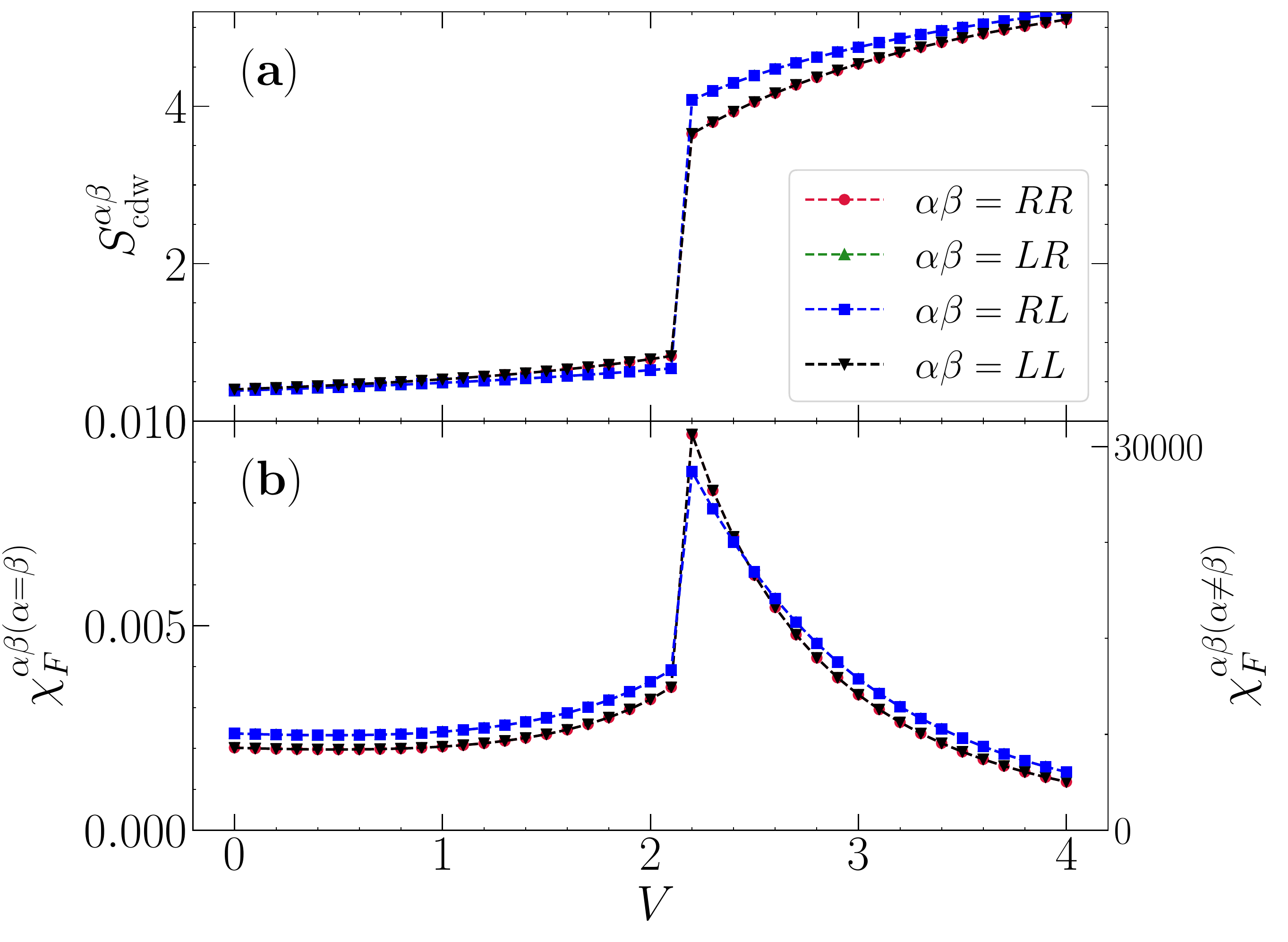}
\caption{(a) The CDW structure factor computed with different combinations of the left and right eigenvectors $|\alpha_L\rangle$ and $|\alpha_R\rangle$. (b) The fidelity susceptibility over the same combinations of left and right states -- note the dual vertical axis. All data are obtained for $\gamma_2=2$.}
\label{fig:app}
\end{figure}

In the main text, we present results for physical quantities when using the right eigenstates, i.e., the ones satisfying $\hat H_{\rm eff} |\alpha_R\rangle = E_\alpha|\alpha_R\rangle$. We notice that one could equally use left eigenvectors, defined as $\hat H^\dagger_{\rm eff} |\alpha_L\rangle = E^*_\alpha|\alpha_L\rangle$. Figure~\ref{fig:app} shows the results for different combinations of left and right eigenvectors for both the CDW structure factor [Fig.~\ref{fig:app}(a)] and the fidelity susceptibility [Fig.~\ref{fig:app}(b)]. The first-order phase transition is clearly captured, resulting in sharp discontinuities for both quantities at the critical interactions. Quantitative differences appear owing to the flexibility in how to mutually define the inner products~\cite{Edvardsson2023} $\langle\alpha_L|\alpha_R\rangle$, $\langle\alpha_L|\alpha_L\rangle$, and $\langle\alpha_R|\alpha_R\rangle$ [our results consider $\langle\alpha_L|\alpha_L\rangle = \langle\alpha_R|\alpha_R\rangle=1$], but qualitatively the results reflect the same physical behavior.

\section{Quantum trajectory method} \label{app:qtm}
Given initial density matrix $\hat \rho(t_{0})=\sum_{\alpha}p_{\alpha}\dyad{\alpha}$, in which $0\leq p_{\alpha}\leq1 $ and $\sum_{\alpha} p_{\alpha}=1$, our objective is to simulate the dynamics of $\rho(t)$ in the time interval $[t_{0}, t_{f}]$. The main part of the quantum trajectory method is
to generate stochastically different evolution paths of wave functions in $[t_{0},t_{f}]$, namely trajectories, and each observable quantity is averaged over all the trajectories to get its expected value. The generation of each trajectory is given by the following algorithm~\cite{Dum1992b, Daley2014}:

\begin{enumerate}
    \item Randomly choose the initial state $|\phi(0) \rangle= |\alpha\rangle$ according to the probabilities $p_{\alpha}$.
    
    \item Generate a random number $r$ from a uniform distribution in $[0,1]$.

    \item Propagate $|\phi(t) \rangle$ according to
    \begin{equation}
        \ii \dv{t}|\phi(t) \rangle= \hat H_{\text{eff}}|\phi(t) \rangle
    \end{equation}
    until time $t^{-}$ where $\norm{\ket{\phi(t^-)}}^{2} = r$.

    \item Calculate weights $w_i = \sqrt{\gamma_i}\norm{\hat L_i \ket{\phi(t^-)}}$
    and randomly choose the jump channel $i$ according to the probabilities $p_{i}
    = w_{i} / \sum_{i} w_{i}$.
    
    \item Calculate the normalized wave function after the quantum jump by
    \begin{equation}
        |\phi(t^{+}) \rangle= \frac{\hat L_{i} |\phi(t^{-}) \rangle}{\norm{\hat L_i \ket{\phi(t^-)}}}\ .
    \end{equation}
    \item Go to 2 and continue the evolution of $|\phi(t) \rangle$ until $t = t_{f}$.
\end{enumerate}

After getting enough independent trajectories, an approximation of the density matrix is given by
\begin{equation}
\label{eq:6}\rho(t) = \biggl\langle\!\!\biggl\langle \frac{\dyad{\phi(t)}}%
{\ip{\phi(t)}{\phi(t)}} \biggr\rangle\!\!\biggr\rangle\ ,
\end{equation}
where $\langle\!\langle\cdot\rangle\!\rangle$ denotes the average over all computed trajectories. In our calculations, we run 200 independent trajectories to extract the average value of the relevant observables ($S_{\rm cdw}$ and total density), resulting in a small standard error for its average value. Furthermore, in step 1, the initial state is taken as the ground state of the Hermitian Hamiltonian [Eq.~\eqref{eq:H_haldane_hubbard}] for the given parameters set, and the time integration in step 3 is performed via a restarted Krylov solver method~\cite{Hernandez2005, Eiermann2006}. Lastly, since step 5 inherently breaks spatial translation invariance, we do not consider this symmetry in building $\hat H_{\rm eff}$ for the dynamics.

\bibliography{refs}
\end{document}